\documentclass[11pt,eps]{article}

\usepackage{graphicx}
\usepackage{epstopdf}
\begin{document}

\begin{center}
{Contribution to the Proceedings of IARD 2006, to appear in {\it Foundations of Physics}}

{Preliminary version dated September 23, 2006, revised October 25, 2006}

\vskip 1cm

{\large { \bf NEUTRINO AND/OR ETHERINO?}}

 {\bf  Ruggero Maria  Santilli}\\
{Institute for  Basic
Research\\ P. O. Box  1577, Palm Harbor,  FL 34682,  U.S.A.}\\
{ ibr@gte.net, http://www.i-b-r.org, 
http://www.neutronstructure.org}
\end{center}

\begin{abstract}
By using a language as accessible to a broad audience as possible, in this paper we review  serious insufficiencies of the hypothesis that neutrinos and quarks are physical particles in our spacetime;   we introduce, apparently  for the first time, the hypothesis that the energy and spin needed for the synthesis of the neutron inside stars originate either from the environment or from the ether conceived as a universal medium with very high energy density via an entity here called {\it etherino,} denoted with the letter "$a$" (from the Latin aether), carrying mass and charge 0, spin 1/2 and $0.78 MeV$ energy according to the synthesis $p^+ + a + e^-\rightarrow n$; we identifies compatibility  $p^+ + a + e^-\rightarrow n\rightarrow p^+ + e^- + \bar \nu$ and incompatibility $p^+ + a + e^-\rightarrow n\rightarrow p^+ + e^- + \bar a$ of the neutrino and etherino hypotheses, the latter representing the possible return of missing features to the ether without being necessarily in conflict with neutrino experiments; we review the  new structure model of the neutron and hadrons at large with massive physical constituents produced free in the spontaneous decays as permitted by the covering hadronic mechanics; we show its compatibility with the standard model when interpreted as only providing the final Mendeleev-type classification of hadrons; we point out basically new clean energies predicted by the new model; we indicate new experiments confirming the above studies although in a preliminary form; and we conclude with the proposal of new resolutory experiments suggested for the much needed search of new clean energies.

\end{abstract}

\begin{center} 

\end{center}

\vskip0.30cm

\noindent {\large {\bf 1. Historical notes.}} 
In 1920, Rutherford [1a] submitted the hypothesis that hydrogen atoms in the
core of stars are compressed into new particles having the size of the proton that he called  {\it neutrons,} according to the synthesis 
$$
p^+ + e^-  \rightarrow n.
\eqno(1.1)
$$
The existence of the neutron was confirmed in 1932 by
Chadwick [1b]. However, Pauli [1c]  noted that the spin 1/2 of the neutron
cannot be represented via a quantum state of two particles each having spin 1/2, and conjectured the possible
emission of a new neutral massless particle with spin 1/2. Fermi [1d] adopted Pauli's conjecture, coined the
name {\it neutrino} (meaning in Italian ``little neutron") with symbol $\nu$ for the particle and $\bar {\nu}$
for the antiparticle,  and
 developed the theory of {\it weak interactions} according to which the synthesis of the neutron is given by 
 $$
p^+ +  e^-  \rightarrow n + \nu,
\eqno(1.2)
$$
with inverse reaction, the spontaneous decay of the neutron, 
$$
n \rightarrow p^+  + e^- + \bar \nu.
\eqno(1.3)
$$
The above hypothesis was more recently incorporated into the so-called {\it standard model} (see, e.g., [1e]) in which the original neutrino was extended to three different particles, the {\it electron, muon and tau neutrinos} and their anti[particles; neutrinos were then assumed to have masses; then to have different masses; and then to "oscillate" (namely, to change "flavor" or transform one type into the other).
\vskip0.30cm


\noindent {\large {\bf 2. Insufficiencies of neutrino hypothesis.}}  
Despite historical advances, the neutrino hypothesis has remained afflicted by a number of basic, although generally unspoken insufficiencies or sheer inconsistencies, that can be summarized as follows:

1) According to the standard model, a neutral particle carrying mass and energy in our spacetime is predicted to cross very large hyperdense media, such as those inside stars, without any collision. Such a view is outside  scientific reason because already questionable when the neutrinos were assumed to be massless. The recent assumption that neutrinos have mass has rendered the view beyond the limit of plausibility because  a massive particle carrying energy in our spacetime simply cannot propagate within  hyperdense media inside large collections of  hadrons without any collision.



\begin{figure}[htbp] 
   \centering
   \includegraphics[width=2in]{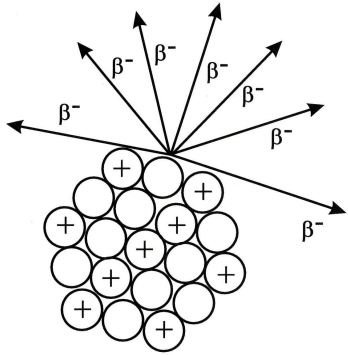} 
 \centering   \caption{{\it  A conceptual illustration of the dependence of the kinetic energy of the electron in nuclear beta decays on the direction of  emission due to the strongly attractive Coulomb interaction between the positively charged nucleus and the negatively charged electron. }}
\end{figure}


2) The fundamental reaction for the production of the (electron) neutrino, Eq. (1.2), violates the principle of conservation of the energy, unless the proton and the electron have kinetic energy of at least $0.78 MeV$, in which case there is no energy available for the neutrino. 
 In fact,  the sum of the rest energies of the proton and the electron ($938.78   MeV$) is $0.78  MeV$ {\it smaller} than the neutron rest energy ($939.56  MeV$).



\begin{figure}[htbp] 
   \centering
   \includegraphics[width=2in]{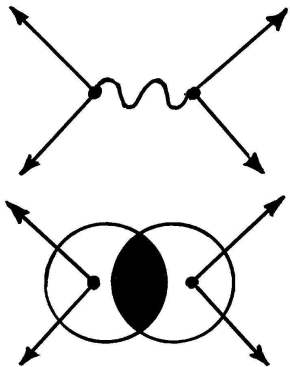} 
 \centering   \caption{{\it  A schematic illustration of the lack of established validity  of the conventional (quantum) scattering theory  in the elaboration of neutrino experiments due to the abstraction of all particles as massive points (top view), while particles used in neutrino experiments are extended with hyperdense medium resulting in deep overlapping and mutual penetrations (bottom view). The latter conditions require a broader scattering theory including nonlocal, nonlinear and non-Hamiltonian effects under which the claimed experimental results are not expected to remain  valid. }}
\end{figure}


3) As reported in nuclear physics textbooks, the energy measured as being carried by the electron
in beta decays is a bell-shaped curve with a maximum value of the order of $0.782 MeV$ (depending on nuclear data). The "missing energy" has been assumed throughout
the 20-th century to be carried by the hypothetical neutrino. However, in view of the strongly attractive Coulomb interactions between the nucleus and the electron, explicit calculations show that the energy carried by the electron depends on the direction of emission, with maximal value for radial emission and minimal value for tangential emission (Figure 1). In this case, the "missing energy" is absorbed by the nucleus, again, without any energy left for the neutrino.

4) The claims of "experimental detection" of neutrinos are perhaps more controversial than the theoretical aspects because of numerous reasons, such as: the lack of established validity of the scattering theory (se Figure 2); the elaboration of the data via a theory centrally dependent on the neutrino hypotheses; the presence in recent "neutrino detectors" of radioactive sources that could themselves account for the extremely few events over an enormous number of total events; the lack of uniqueness of the neutrino interpretation of experimental data due to the existence of alternative interpretations without the neutrino hypothesis; and other aspects.

5) Numerous additional insufficiencies exist, such as: the absence of  well identified physical differentiations between the electron, muon and tau neutrino; the theory contains an excessive number of parameters essentially capable to achiever any desired fit; the probability of the synthesis of the neutron according to Eq. (1.2) is virtually null when the proton and the electron have the needed threshold energy of $0.78 MeV$ due to their very small scattering cross section ($10^{-20}$ barns) at the indicated energy; particles are characterized by positive energies and antiparticles by negative energies, with the consequential lack of plausibility of the conjugation from neutrino to antineutrino in the transition from Eq. (1.2) to (1.3) since the same conjugation does not exist for the proton and the electron; and other insufficiencies, a compatibility condition that would require the reaction $n\rightarrow p^+ + e^- + \nu$ (rather than $\bar \nu$).

For additional studies on the insufficiency os sheer inconsistencies of the neutrino hypothesis, one may consult Bagge [2a] and Franklin [2b] for an alternative theories without the neutrino hypothesis; Wilhelm [2c] for additional problematic aspects; Moessbauer [2d] for problems in neutrino oscillations; Fanchi [2e] for serious biases in "neutrino experiments"; and literature quoted therein.

On historical grounds it should be noted that the original calculations in beta decays were done in the 1940s via {\it the abstraction of nuclei as massive points} because mandated by the axioms of quantum mechanics, in which case indeed there is no dependence of the electron-nucleus Coulomb interaction on the direction of beta emission, and the neutrino hypothesis becomes necessary.

However, the abstraction of nuclei as massive points nowadays implies a violation of scientific ethics and accountability since nuclei are very large objects for particle standards. The representation of nuclei as they actually are in the physical reality then requires the abandonment of the neutrino conjectures in favor of more adequate  vistas.

\vskip0.30cm


\noindent {\large {\bf 3. Insufficiencies of quark hypothesis.}}  Some of the fundamental, yet generally unspoken insufficiencies or sheer inconsistencies of the assumption that quarks are physical particles in our spacetime are the following:

1) According to the standard model [5], at the time of the synthesis of the neutron according to Eq. (1.2), the proton and the electron literally "disappear" from the universe to be replaced by hypothetical quarks as neutron constituents. Moreover, at the time of the neutron spontaneous decay, Eq. (1.3), the proton and the electron literally "reappear" again. This view is beyond 
 scientific reason, because the proton and the electron are the only {\it permanently stable} massive  particles clearly established so far and, as such, they simply cannot "disappear" and then "reappear"  because so desired by quark supporters.   The {\it only} plausible hypothesis under Eqs. (1.2) and (1.3) is that the proton and the electron are actual physical constituents of the neutron as originally conjectured by Rutherford, although the latter view requires the adaptation of the theory to physical reality, rather than the opposite attitude implemented by quark theories.

2) When interpreted as physical particles in our spacetime, {\it quarks cannot experience any gravity.} As clearly stated by Albert Einstein in his limpid writings, gravity can only be defined in spacetime, while quarks can only be defined in the  mathematical, internal, complex valued unitary space with no possible connection to our spacetime (because prohibited by the O'Rafearthaigh's theorem). Consequently, physicists who support the hypothesis that quarks are the physical constituents of protons and neutrons, thus of all nuclei, should see their body levitate due to the absence of gravity.

3) When, again, interpreted as physical particles in our spacetime, {\it quarks cannot have any inertia.} In fact, inertia can only be rigorously admitted for the eigenvalues of the second order Casimir invariant of the Poincar\'e symmetry, while quarks cannot be defined via such a basic spacetime symmetry, as expected to be known by experts to qualify as such. Consequently, "quark masses" are purely mathematical parameters deprived of technical characterization as masses in our spacetime.

4) Even assuming that, with unknown scientific manipulations, the above inconsistencies are
resolved, it is known by experts that quark theories have failed to achieve a representation of {\it all} 
characteristics of hadrons, with catastrophic insufficiencies in the representation of spin,
magnetic moment, mean lives, charge radii and other basic features of hadrons.

5) It is also known by experts that the application of quark conjectures to the structure of nuclei has
multiplied the controversies, while resolving none of them. As an example, the assumption
that quarks are the constituents of protons and neutrons in nuclei has failed to achieve a
representation of the main characteristics of the simplest possible nucleus, the deuteron. In fact, quark
conjectures are unable to represent the spin 1 of
the deuteron (since they predict spin zero in the ground state of two particles each having spin ${1\over 2}$), they are
unable to represent the anomalous magnetic moment of the deuteron despite all possible relativistic corrections attempted for decades (because the presumed quark orbits are too small to fit data following polarizations or deformations), they are unable to represent the 
stability of the neujtron when a deuteron constituent, they are unable to represent the charge radius of the deuteron, and when passing to larger nuclei,
such as the zirconium, the catastrophic inconsistencies of quark conjectures can only be defined as being
embarrassing.

For additional references, one may consult Ref. [3a] on historical reasons preventing quarks to be physical particles in our spacetime; Ref. [3b] on a technical treatment of the impossibility for quarks to have gravity or inertia; Ref. [3c]
on a more detailed presentation on the topic of this section; and Refs. [7,9g,9h] for general studies.

The position  adopted by the author since the birth of quark theories (see the memoir [3a] of 1981), that appears to be even more valid nowadays, 
is that  {\it the unitary, Mendeleev-type,  SU(3)-color classification of particles 
into families has a final character. Quarks are purely mathematical representation of a purely mathematical internal symmetry solely definable on a purely mathematical, complex-valued unitary space. As such, the use of quarks is indeed necessary for the elaboration of the theory, as historically suggested by the originator  Murray Gell-Mann, but quarks are not physical particles in our spacetime. }

Consequently, the identification of the hadronic constituents with physical particles truly existing in our spacetime is more open than ever and carries ever increasing societal implications since the assumption that quarks are physical constituents of hadrons prevents due scientific process  in the search for new clean energies so much needed by mankind, as illustrated later on.

Needless to say, all alternative structure models, including those without neutrino and
quark conjectures must achieve full compatibility with said unitary models of classification, in 
essentially the same way according to which quantum structures of atoms achieved full
compatibility with their Mendeleev classification.

On historical grounds, the classification of nuclei, atoms and molecules required {\it two different
models,} one for the classification and a separate one for the structure of the individual element of a
given family. Quark theories depart from this historical teaching because of  their conception of representing with one single theory both the classification and the structure of hadrons. 

The view advocated in this paper is that, quite likely, history will repeat itself. The transition from the Mendeleev classification of atoms to the atomic structure required a basically new theory, quantum mechanics. Similarly, the transition from the Mendeleev-type classification of hadrons to the structure of individual hadrons will require a broadening of the basic theory, this time a generalization of quantum mechanics due tg the truly dramatic differences of the dynamics of particles moving in vacuum, as in the atomic structure, to the dynamics of particles moving within hyperdense media as in the hadronic structure.

\vskip0.30cm


\noindent {\large {\bf 4. Inapplicability of quantum mechanics for the synthesis and structure of  hadrons.}} Pauli, Fermi, Schr\"odinger and other founders of quantum mechanics pointed out that  synthesis of the neutron according to Rutherford  is impossible for the following reasons:

1) As indicated in Section 1, quantum mechanics cannot represent the ${1\over 2}$ spin of the neutron according to Rutherford's conception because 
the total angular momentum of a ground state of a two particles with spin 1/2, such as the proton and
the electron, must be 0.

2) The representation of synthesis (1.1) via
quantum mechanics is impossible because it would require a "positive" binding-like energy, in violation of basic
quantum laws requiring that all binding energies must be negative, as fully established for nuclei, atoms and molecules. This is
due to the fact indicated in Section 2 that the sum of the mass of the proton and of the electron,
$$
 m_p + m_e  = 938.272 Mev +  0.511 MeV = 938.783 MeV,
\eqno(4.1)
$$ 
is {\it smaller} than the mass of the neutron, $m_n = 939.565 MeV$, with "positive" mass defect
$$
m_n - (m_p + m_e) =  939.565 - (938.272 +  0.511) MeV = 0.782 MeV.
\eqno(4.2)
$$
Under these conditions all quantum equations
become physically inconsistent in the sense that mathematical solutions are indeed admitted, but the indicial equation of Schr\"odinger's equation no longer admits the representation of the total energy and other physical quantities with real numbers (readers seriously interested in the synthesis of hadrons are strongly suggested to attempt   the solution of any quantum bound state in which the conventional negative binding energy is turned into a positive value).

3) Via the use of the magnetic moment of the proton $\mu_p = 2.792 \mu_N$ and of the electron $\mu_e = 1.001
\mu_B$, it is impossible to reach the magnetic moment of the neutron $\mu_n = - 1.913 \mu_N$.

4) When the neutron is interpreted as a bound state of one proton and one electron, it is impossible to
reach the neutron meanlife $\tau_n = 918 sec$ that is quite large for particle standards, since quantum
mechanics would predict the expulsion of the electron in nanoseconds.

5) There is no possibility for quantum ,mechanics to represent the neutron charge radius of about $1 F = 10^{-13}
cm$ since the smallest predicted radius is that of the hydrogen atom of $10^{-8} cm$, namely 5,000 times
{\it bigger} than that of the neutron.



\begin{figure}[htbp] 
   \centering
   \includegraphics[width=2in]{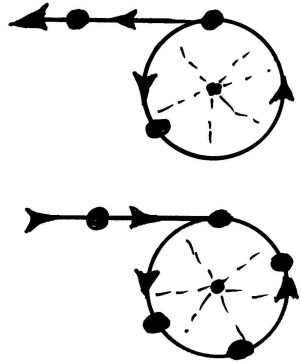} 
 \centering   \caption{{\it A schematic view of the transformation of linear into angular motion and vice-versa that could play a crucial role in the synthesis of the neutron and its stimulated decay. Note that such a transformation is outside the capabilities of the Poincar\'e symmetry due to its sole validity  for Keplerian systems, that is, for massive points orbiting around a heavier nucleus without collisions. By comparison, the transformation of this figure requires the presence of subsidiary constraints  altering the conservation laws, thus altering the very generator of the applicable symmetry. }}
\end{figure}


It should be noted that the above insufficiencies of quantum mechanics generally apply for the synthesis of all hadrons, beginning with that for the neutral pion
$$
e^+ + e^- \rightarrow \pi^o,
\eqno(4.3)
$$
where the "positive binding energy" is now of $133.95 MeV$.

The view advocated in this paper is that, rather than denying the synthesis of hadrons just because not permitted by quantum mechanics, a covering mechanics  permitting quantitative studies of said synthesis should be built.

The most visible evidence indicating the lack of exact character of quantum mechanics for the synthesis and structure of hadrons is that, unlike atoms, {\it hadrons do not have nuclei.} Consequently, a mechanics that is exact for the atomic structure cannot be exact for the hadronic structure due to the lack of a Keplerian structure that, in turn, requires the breaking of the fundamental Galilean and Loretzian symmetries.

Quantum mechanics was conceived and constructed for the representation of the trajectories of electrons
moving in vacuum in atomic orbits (this is the so-called {\it exterior dynamical problems}), in which case the theory received historical verifications. 
The same mechanics cannot possibly be exact for the description of the dramatically different physical
conditions of the same electron moving within the hyperdense medium inside a proton (this is the so-cal;led {\it interior dynamical problems}). Such an
assumption literally implies the belief in the perpetual motion within a physical medium since it implies
that an electron must orbit in the core of a star with a conserved angular momentum, as requested by the
quantum axiom of the rotational symmetry and angular momentum conservation law.

In the final analysis it has been established by scientific history that the validity of any given theory
within given conditions is set by the results. Quantum mechanics has represented {\it all} features of the
hydrogen atom in a majestic way and, therefore, the theory is exactly valid for the indicated
conditions. 

By contrast, when extended to the structure of particles, quantum mechanics has only produced
an interlocked chain of individually implausible and unverifiable conjectures on neutrinos and quarks while having dramatic insufficiencies in the representation of particle data, besides failing to
achieve final results in various other branches of sciences, such as in nuclear physics, chemistry and
astrophysics.  

After all these controversies protracted for such a long
period of time, there comes a time in which the serious conduction of serious science requires a
re-examination of the foundational theory.

\vskip0.30cm


\noindent {\large {\bf 5. The etherino hypothesis.}}  As clearly shown by the preceding analysis, the synthesis of the neutron according to Rutherford [1a] not only misses spin ${1\over 2}$ as historically pointed out by Pauli [1c] and Fermi [1d], but also misses $0.78 MeV$ energy. Moreover, these quantities must be {\it acquired} by the proton and electron for the synthesis to exist, rather than being "released" as in Eq. (1.2).

Consequently, a central open problem in the synthesis of the neutron is the identification of "were" these quantities originate. The first evident answer is that the missing quantities originate from the environment in the interior of stars in which the neutron is synthesized. In fact, there is no doubt that the interior of stars can indeed supply spin ${1\over 2}$ and $0.78 MeV$ energy.

However, due to the fundamental character of the neutron synthesis for the entire universe, serious  studies should not be solely restricted  to the most obvious possibility, and should consider instead all plausible alternatives no matter how speculative they may appear at this time.

Along the latter lines, we recall the hypothesis of the continuous creation of matter in the universe that has been voiced repeatedly during the 20-th century. 
In this paper we point out that the best possible mechanism for continuous creation is precisely the synthesis of neutrons inside stars under the assumption that the missing energy and spin originates from the ether conceived as a universal medium with an extremely large energy density.

Far from being farfetched, the hypothesis is supported by predictably insufficient, yet significant evidence, such as the fact that stars initiate their lives as being solely composed of hydrogen atoms that miss the energy and spin needed for the first synthesis, that of the neutron, after which all conventional nuclear syntheses follow.

Additionally, explicit calculations indicate that the immense energy needed for a supernova explosion simply cannot be explained via the sole use of conventional nuclear syntheses, thus suggesting again the possible existence of a mechanism extracting energy from the ether and transforming it into a form existing in our spacetime.

It is important to point out that the notion of ether as a universal substratum appears to be necessary not only for the characterization and propagation of electromagnetic waves, but also for the characterization and propagation of all elementary particles and, therefore, for all matter existing in the universe.

The need for
a universal medium for the characterization and propagation of electromagnetic ``waves" is so strong to require no study here, e.g., for waves with $1 m$ wavelength for which the reduction to photons for the purpose of eliminating the ether loses credibility.

The same notion of ether  appears necessary also for the characterization and propagation of the electron, due to its structure as  a "pure oscillation" of the ether, namely, an oscillation of one of its points without any oscillating mass as conventionally understood. Similar structures are expected for all other truly elementary particles.

It should be indicate that the above conception implies that, contrary to our sensory perception, {\it matter is totally empty and space is totally full,} as suggested by the author since his high school studies [4a].  This conception is necessary to avoid the ``ethereal
wind" [4b] that delayed studies on the ether for at least one century, since motion of matter would merely require the transfer of
the characteristic oscillations from given points of the ether to others. Mass is then characterized by the known equivalence of the energy of the characteristic oscillations, and inertia is the resistance provided by the ether against changes of motion. For additional recent views on the ether we refer interested readers to Ref. [4c].

In order to conduct quantitative studies of the above alternatives, in this note we submit the hypothesis, apparently for the first time, that the synthesis of the neutron from protons and electrons occurs via the absorption either from the environment inside stars or from the ether of an entity, here called {\it etherino} (meaning in Italian "little ether") and indicated with the symbol "$a$" (from the latin aether) having mass and charge 0, spin 1/2 and  a minimum of $0.78 MeV$ energy. We reach in this way the following

{\bf Etherino hypothesis on the neutron synthesis}
 $$
 p^+ + a_n + e^-  \rightarrow  n,
 \eqno(5.1)
 $$
 where $a_n$ denotes the {\it neutron etherino} (see below for other cases), and the energy $0.78 MeV$ is assumed to be ''minimal'' because of the conventional  "negative" binding energy due to the attractive Coulomb interactions between the proton and the electron at short distances.
 
 The apparent {\it necessity} of the etherino hypothesis is due to the fact that the use of an antineutrino in lieu of the etherino
 $$
 p^+ + \hat \nu + e^-  \rightarrow  n,
 \eqno(5.2)
 $$
would have no known physical value for various reasons, such as: 1) the proton and/or the electron cannot possibly absorb $0.78 MeV$ energy and spin ${1\over 2}$ from the antineutrino due to the virtually null value of their cross section; 2) being an antiparticle, the antineutrino has to carry a {\it negative} energy [7c], while the synthesis of the neutron requires a {\it positive} energy; and others.

In the author view, a compelling aspect supporting the etherino hypothesis is the fact that the synthesis of the neutron has the highest probability when the proton and the electron are at relative rest, while the same probability becomes essentially null when the proton and the electron have the (relative) missing energy of $0.78 MeV$ since in that case their cross section becomes very small (about $10^{-20} barns$).

Another argument supporting the etherino over the neutrino hypothesis is that the former permits quantitative studies on the synthesis of the neutron as we shall see in subsequent sections, while the latter provides none, as shown in the preceding section. 

Still another  supporting argument is that the etherino hypothesis eliminates the implausible belief that  massive particles carrying energy in our spacetime can traverse enormous hyperdense media without collisions  since the corresponding etherino event could occur in the ether as a universal substratum, without any propagation of mass and energy in our spacetime.

In order to prevent the invention of additional hypothetical particles over an already excessive number of undetected particles existing in contemporary physics, the author would like to stress that {\it the etherino is not intended to be a conventional particle existing in our spacetime, but an entity representing the transfer of the missing quantities from the environment or the ether to the neutron.} The  lack of characterization as a conventional physical particle will be made mathematically clear in the next sections.
 
 It is evident that the etherino hypothesis requires a reinspection of the spontaneous decay of the neutron. To conduct a true scientific analysis, rather than adopt a scientific religion, it is necessary to identify all plausible alternatives, and then reach a final selections via experiments. We reach in this way the following {\it three} possible alternatives:
 
 {\bf First hypothesis on the neutron decay:}
 $$
 p^+ + a_n + e^-  \rightarrow  n\rightarrow p^+ + e^- + \hat \nu,
 \eqno(5.3)
 $$
 namely, the etherino hypothesis for the neutron  "synthesis" can indeed be fully compatible with the neutrino hypothesis for the neutron "decay";
 
  {\bf Second hypothesis on the neutron decay:}
 $$
 p^+ + a_n + e^-  \rightarrow  n\rightarrow p^+ + e^- + a_n,
 \eqno(5.4)
 $$
 namely, we have the return of the missing energy and spin to the environment or the ether; and
 
   {\bf Third hypothesis on the neutron decay:}
 $$
 p^+ + a_n + e^-  \rightarrow  n\rightarrow p^+ + e^-,
 \eqno(5.5)
 $$
 namely, no neutrino or etherino is emitted. Note that the latter case is strictly prohibited by quantum mechanics because of known conservation laws and related symmetries.  However, the latter case should be not dismissed superficially because it is indeed admitted by the covering hadronic mechanics via the conversion of the orbital into the kinetic motion as in Fig. 3.

The synthesis of the antineutron in the interior of antimatter stars is evidently given by
$$
p^- + \bar {a}_{\bar n} +  e^+ \rightarrow \bar {n}.
\eqno(5.6)
$$
where $\bar a_{\bar n}$ is the {\it antineutron antietherino,} namely an entity carrying negative energy as apparently necessary for antimatter [7c]. This would imply that the ether is constituted by a superposition of very large but equal densities of positive and negative energies existing in different yet coexisting spacetimes, a concept with even deeper cosmological and epistemological implications since their total null value would avoid  discontinuities at creation.

For the synthesis of the neutral pion we have the hypothesis
$$
e^+ + a_{\pi^o} + e^- \rightarrow \pi^o,
\eqno(5.7)
$$
where $a_{\pi^o}$ is the {\it $\pi^o$-etherino,} namely, an entity carrying mass, charge and spin 0 and minimal energy of $133.95 MeV$  transferred from the ether to our spacetime. Additional forms of etherino can be  formulated depending on the synthesis at hand.

The understanding of synthesis (5.7) requires advanced knowledge of modern classical and operator theories of antimatter (see monograph [7c]) because $a_{\pi^o}$ must be {\it iso-self-dual,} namely, it must coincide with its antiparticle as it is the case for the $\pi^o$. In more understandable terms, $a_{\pi^o}$ represents an equal amount of positive and negative energy, since only the former (latter) can be acquired by the electron (positron), the sign of the total energy for isoselfdual states being that of the observer [{\it loc. cit.}]. 

Intriguingly, the etherino hypothesis for the neutron {\it decay,} Eq. (5.4), is not necessarily in conflict with available data on neutrino experiments, because it could provide their mere re-interpretation as a new form of communication through the ether. Moreover, in the event the propagation of the latter event results to be longitudinal as expected, it could be much faster than  the speed of conventional (transversal) electromagnetic waves.  

In the final analysis, the reader should not forget that, when inspected at interstellar or intergalactic distances, communications via electromagnetic waves should be compared to the communications by the Indians with smoke signals. The search for basically new communications much faster than those via electromagnetic waves is then mandatory for serious astrophysical advances. In turn, such a search can be best done via longitudinal signals propagating through the ether. Then, the possibility of new communications being triggered by the etherino reinterpretation of the neutrino should not be aprioristically dismissed without serious study.

\vskip0.30cm


\noindent{\large {\bf 6. Rudiments of the covering hadronic mechanics.}}   When at the Department of Mathematics of
Harvard University in the late 1970s, R. M. Santilli [5a] proposed the construction of a new broader realization of the axioms of quantum mechanics under the name of {\it hadronic mechanics} that was intended for the  solution of the insufficiencies of  conventional theories   outlined
in the preceding sections. The name  "hadronic " mechanics" was selected to emphasize the primary applicability of the new mechanics at the range of the strong interactions, since the validity of quantum mechanics for bigger distances was assumed {\it a priori.} .

The central problem was  to identify a broadening-generalization of quantum
mechanics  in such a way to represent  linear,
local and potential interactions, as well as additional, contact,
nonlinear, nonlocal-integral and nonpotential interactions, as expected in the neutron synthesis. as well as in deep mutual penetration and overlapping of hadrons (Figure 2).

Since the Hamiltonian can only represent conventional local-potential interactions, the
above condition requested the identification of a {\it  new quantity} capable
of representing  interactions that, by conception, are outside the capability of
a Hamiltonian. Another necessary condition was the exiting from the class of
equivalence of quantum mechanics, as a consequence of  which the broader theory
had to be {\it nonunitary,} namely, its time evolution had to violate the
unitarity condition.  The third and most insidious condition was the {\it time 
invariance,} namely, the broader mechanics had to predict the same numerical values under the same conditions at different times. 

It was evident that a solution verifying the above conditions required {\it
new mathematics, e.g. new numbers, new spaces, new geometries, new symmetries,
etc.} A detailed search in advanced mathematical libraries of the
Cantabridgean area revealed that the needed new mathematics simply did not exist and, therefore, had to be built. 

Following a number of (unpublished) trials and errors, Santilli [5a] proposed 
the solution consisting in the representation
of  contact, nonlinear, nonlocal and nonpotential interactions via a {\it
generalization (called lifting) of the basic unit} $+1$ of quantum mechanics into a
function, a matrix or an operator $\hat I$ that is positive-definite like $+1$, but
otherwise has an arbitrary functional dependence on all needed quantities, such
as time $t$, coordinates $r$, momenta
$p$, density $\mu$, frequency $\omega$, wavefunctions $\psi$, their derivatives
$\partial \psi$, etc.
$$
+1 > 0 \; \; \; \rightarrow \; \; \; \hat I(t, r, p, \mu. \omega, \psi. \partial\psi,
...) = \hat I^\dag =  1 / \hat T > 0,
\eqno(6.1)
$$
while jointly lifting the conventional associative product
$\times$ between two generic quantities $A, B$ (numbers, vector fields, matrices,
operators, etc.) into the form admitting
$\hat I$, and {\it no longer}
$+1$, as the correct left and right unit
$$
A\times B \; \; \; \rightarrow \; \; \; A\hat \times B = A\times \hat T\times B,
\eqno(6.2a)
$$
$$
1\times A = A\times 1 = A \; \; \; \rightarrow \; \; \; \hat I\hat \times A = A\hat
\times \hat I = A,
\eqno(6.2b)
$$
for all elements $A, B$ of the set considered.

The selection of the basic unit resulted to be unique for the verification of the
above three conditions. As an illustration, whether generalized or not, the unit is
the basic invariant of any theory. The representation of non-Hamiltonian
interactions with the basic unit permitted the crucial by-passing of the
theorems of catastrophic inconsistencies of nonunitary theories (for a review one may inspect Section 5, Chapter 1 of Ref. [7a]).
Since the unit is the ultimate pillar of all mathematical and physical
formulations, liftings (6.1) and (6.2) requested a corresponding, compatible
lifting of the {\it totality} of the mathematical and physical formulations used by
quantum mechanics, resulting indeed into new numbers, new
fields, new spaces, new algebras, new geometries, new symmetries, etc, [5b,5c].
Mathematical maturity in the formulation of
the new numbers was reached only in memoir [5b] of 1993 and general mathematical
maturity was reached in memoir [5c] of 1996. Physical maturity was then quickly
achieved in papers [5d-5f].

The fundamental dynamical equations of hadronic mechanics were submitted by Santilli
in the original proposal [5a], are today called {\it Heisenberg-Santilli
isoequations,} and can be written in the finite form

$$
\hat A(\hat t) = \hat U(\hat t)\hat \times \hat A(\hat 0)\hat \times \hat
U^\dag(\hat t) =
(\hat e^{ \hat H\hat \times  \hat t\hat \times \hat i})\hat \times \hat A(\hat 0)
\hat
\times (\hat e^{-\hat i\hat \times \hat t\hat \times \hat H}) =
$$
$$
= [(e^{H\times \hat T\times t\times i})\times \hat I]\times \hat T\times A(0) \times
\hat T\times [\hat I\times (e^{-i\times t\times \hat T \times H})] =
$$
$$
(e^{H\times \hat T\times t\times i})\times \hat A(\hat 0) \times (e^{-i\times t\times \hat
T \times H}),
\eqno(6.3a)
$$
$$
\hat U = \hat e^{i\times H\times t}, \; \; \;  \hat U^\dag = \hat e^{-i\times t\times \times H}, \; \; 
\hat U\hat \times \hat U^\dag = \hat U\hat \times \hat U = \hat I \not = 1, \; \; [H. \hat T] \not = 0,
\eqno(6.3b)
$$
and infinitesimal form 
$$
\hat i\hat {\times} {\hat d\hat A\over \hat d\hat t} =
i\times \hat I_t\times {d\hat A\over d\hat t} = [\hat A\hat ,
\hat H] = \hat A\hat {\times}\hat H \hat - \hat H\hat {\times}\hat A =
$$
$$
= \hat A\times \hat T(\hat t, \hat r, \hat p, \hat \psi, \hat
\partial \hat \psi, ...)\times \hat H - \hat H\times \hat T(\hat t,
\hat r, \hat p, \hat \psi, \hat
\partial \hat \psi,
...)\times \hat A,
\eqno(6.4)
$$
where: Eq. (6.3b) represent the crucial {\it nonunitarity-isounitary property,} namely, the violation of unitarity on conventional Hilbert spaces over a field, and its reconstruction on  {\it iso-Hilbert spaces} over {\it isofields} with inner
product $<\hat \psi|\hat \times |\hat \psi>$; we have used in Eqs. (6.3b) the notion of {\it isoexponentiation,} see Eq. (6.12d); all quantities with a ''hat'' are formulated  on isospaces over isofields with isocomplex numbers $\hat c =
c\times \hat I$, $c\in C$; and one should note the {\it isodifferential calculus} with
expressions of the type $\hat d /\hat d\hat t = \hat I_t\times d / d\hat t$ first
achieved in memoir [5c].

The equivalent lifting of Schr\"odinger's equation was  suggested by Santilli and other authors  over conventional fields, thus violating the condition of time invariance (see [7a] for historical notes and quotations). The final version was reached by Santilli in memoir [5c] following the construction of the {\it isodifferential calculus} and can be written
$$
\hat i\hat {\times} {\hat {\partial}\over \hat {\partial}\hat
t}|\hat {\psi}> = i \times \hat I_t\times {\partial\over \partial \hat t} |\hat \psi> =
\hat H\hat {\times} |\hat {\psi}> =
$$
$$
= \hat H(\hat t, \hat r, \hat p)\times \hat T(\hat r, \hat p, \hat
{\psi}, \hat {\partial}\hat {\psi},. ...)\times |\hat {\psi}> = \hat
E\hat {\times} |\hat {\psi}> = E\times |\hat \psi>,
\eqno(6.5a)
$$
$$
 \hat p_k\hat {\times}|\hat
{\psi}> = - \hat i\hat {\times}\hat {\partial}_k|\hat {\psi}> = -
i\times \hat I_k^i\times \partial_i|\hat {\psi}>,
\eqno(6.5b)
$$
with {\it isocanonical commutation rules}
$$
[\hat r^i\hat ,  \hat p_j] = \hat i\hat {\times}\hat {\delta}^i_j =
i\times \delta^i_j\times \hat I, [\hat r^i, \hat r^j] = [\hat p_i,
\hat p_j] = 0.
\eqno (6.6)
$$

{\it isoexpectation values}
$$
<\hat A> = {<\hat \psi|\hat \times \hat A\hat\times |\hat \psi>\over <\hat \psi|\hat
\times |\hat \psi>}
\eqno(6.7)
$$
and basic properties
$$
{<\hat \psi|\hat \times \hat I\hat\times |\hat \psi>\over <\hat \psi|\hat \times
|\hat \psi>} = \hat I,\; \; \hat I\hat {\times} |\hat {\psi}> = |\hat {\psi}>, \; \; \; 
\hat I^{\hat n} = \hat I\hat \times \hat I\hat \times ... \hat I \equiv \hat I, \; \; \;
\hat I^{\hat {1/2}} = \hat I,
\eqno (6.8)
$$
the latter confirming that $\hat I$ is indeed the isounit of hadronic mechanics (where the
isoquotient $\hat / = /\times \hat I$ has been tacitly used [5c]).

A few comments are now in order. In honor of Einstein's vision on the lack of completion of quantum mechanics,
Santilli submitted  the original Eqs. (6.1)-(6.8) under the name of {\it isotopies}, a word used in the Greek meaning of ''preserving the original
axioms.'' In fact, $\hat I$ preserves all topological properties of $+1$, $A\hat
\times B$ is as associative as  the conventional product $A\times B$ and the
preservation of the original axioms holds at all subsequent levels to such an extent
that, in the event any original axiom is not preserved, the
lifting is not isotopic. Nowadays, the resulting new mathematics is known as {\it
Santilli isomathematics}, $\hat I$ is called {\it Santilli's isounit},
$A\hat \times B$ is called the {\it isoproduct,} etc. (see  the General Bibliography of Ref. [7a] and monograph [8]).

Note the identity of Hermiticity and its
isotopic image, $(<\hat \psi|\hat \times \hat H^{\hat \dag})\hat \times |\hat \psi> \equiv
<\hat \psi|\hat \times (\hat H\hat \times |\hat \psi>),  \hat H^{\hat \dag} \equiv \hat
H^\dag$, thus implying that all quantities that are observable for quantum mechanics
remain observable for hadronic mechanics; the new  mechanics is indeed isounitary, thus
avoiding the theorems of catastrophic inconsistencies of nonunitary theories; hadronic
mechanics preserves all conventional quantum laws, such as Heisenberg's uncertainties,
Pauli's exclusion principle, etc.; dynamical equations (6.3)-(6.8) have  been
proved to be ''directly universal'' for all
possible theories with conserved total energy, that is, capable of representing all infinitely
possible systems of the class admitted (universality) directly in the frame of the observer without the use of 
transformations (direct universality); and numerous other features one can study in
Refs. [6-8].

Also, one should note that {\it hadronic mechanics verifies the abstract axioms
of quantum mechanics to such an extent that the two mechanics coincide at the abstract,
realization-free level.} In reality, hadronic mechanics provides an explicit and
concrete realization of the theory of ''hidden variables'' $\lambda$, as one can
see from the abstract identity of the isoeigenvalue equation $\hat H\hat \times |\hat
\psi> = \hat E\hat \times |\hat \psi>$ and the conventional  equation
$H\times |\psi> = E\times |\psi>$, by providing in this way an {\it operator}
realization of hidden variables $\lambda = \hat T$ (for detailed studies on these aspects,
including the {\it inapplicability} of Bell's inequality, see Ref. [6g].

We should also indicate that the birth of hadronic mechanics can be seen in the following
{\it new isosymmetry,} here expressed for a constant $K$ for simplicity,
$$
<\psi|\times |\psi>\times 1 \equiv <\psi|\times K^{-1}\times |\psi>\times (K\times
1) = <\psi|\hat \times |\psi>\times \hat I.
\eqno(6.9)
$$

The reader should not be surprised that the above  isosymmetry
remained unknown throughout the 20-th century. In fact, its identification required
the prior discovery of {\it new numbers,} Santilli's isonumbers with
arbitrary units [5b].

Compatibility between hadronic and quantum mechanics is reached via the condition
$$
Lim_{r>>10^{-13} cm} \hat I \equiv 1,
\eqno(6.10)
$$
under which hadronic mechanics recovers quantum mechanics uniquely and identically
at all levels. Therefore, hadronic mechanics coincides with quantum mechanics everywhere except in the interior of the so-called {\it hadronic horizon} (a
sphere of radius $1 F = 10^{-13} cm$) in which the new mechanics admits non-Hamiltonian realizations of strong interactions.

A simple method has been identified in Refs. [5d] for the
construction of hadronic mechanics and all its underlying new mathematics consisting of:

(i) Representing all conventional interactions with a Hamiltonian
$H$ and all non-Hamiltonian interactions and effects with the
isounit $\hat I$;

(ii) Identifying the latter interactions with a nonunitary transform
$$
U\times U^{\dagger} = \hat I \not = I
\eqno(6.11)
$$
(iii) Subjecting the {\it totality} of conventional mathematical, physical
and chemical quantities and all their operations to the above nonunitary
transform, resulting in expressions of the type
$$
I\rightarrow \hat I = U\times I\times U^{\dagger} = 1/\hat T,
\eqno(6.12a)
$$
$$
a\rightarrow \hat a = U\times a\times U^{\dagger} = a\times \hat I,
\eqno(6.12b)
$$
$$
a\times b\rightarrow U\times (a\times b)\times U^{\dagger} =
$$
$$
= (U\times a\times U^{\dagger})\times (U\times
U^{\dagger})^{-1}\times (U\times b\times U^{\dagger}) = \hat a\hat
{\times}\hat b,
\eqno(6.12c)
$$
$$
e^A\rightarrow  U\times e^A\times U^{\dagger} = \hat I\times e^{\hat
T\times \hat A} = (e^{\hat A\times \hat T})\times \hat I,
\eqno(6.12d)
$$
$$[X_i, X_j]\rightarrow U\times [X_i
 X_j]\times U^\dagger =
$$
$$
=  [\hat X_i\hat ,\hat X_j] = U\times (C_{oj}^k\times X_k)\times
U^{\dagger} = \hat C_{ij}^k\hat {\times}\hat X_k =
= C_{ij}^k\times \hat X_k,
\eqno(6.12e)
$$
$$
<\psi | \times |\psi >\rightarrow U\times <\psi | \times |\psi
>\times U^{\dagger} =
$$
$$
= <\psi | \times U^{\dagger}\times (U\times U^{\dagger})^{-1}\times
U\times |\psi >\times (U\times U^{\dagger}) 
= <\hat \psi |\hat {\times} |\hat \psi >\times \hat I,
\eqno(6.12f)
$$
$$
H\times |\psi>\rightarrow U\times (H\times |\psi>) = (U\times
H\times U^{\dagger})\times (U\times U^{\dagger})^{-1}\times (U\times
|\psi>) =
$$
$$
= \hat H\hat {\times} |\hat {\psi}>, etc.
\eqno (6.12g)
$$

Note  that {\it catastrophic inconsistencies emerge in the event even one single
quantity or operation is not subjected to isotopies.} In the absence of
comprehensive liftings, we would have a situation equivalent to the elaboration of the
quantum spectral data of the hydrogen atom with isomathematics, resulting of dramatic
deviations from reality.

It is easy to see that the application of an additional nonunitary
transform  to expressions (6.12) causes the {\it lack of
invariance,} e.g.,
$$
W\times W^\dag \not = I, \; \; \; I\rightarrow \hat I' = W\times \hat I\times W^\dag \not = \hat I,
\eqno(6.13)
$$
with consequential activation of the theorems of catastrophic
inconsistencies [7a]. However, any given nonunitary transform can be identically
rewritten in the isounitary form,
$$
W\times W^{\dagger} = \hat I,\; \; \;  W = \hat W\times \hat T^{1/2},
\eqno(6.14a)
$$
$$
W\times W^{\dagger} = \hat W\hat {\times}\hat W^{\dagger} = \hat
W^{\dagger}\hat {\times}\hat W = \hat I, 
\eqno(6.14b)
$$
 under which hadronic mechanics is indeed isoinvariant
$$
\hat I\rightarrow \hat I' = \hat W\hat {\times}\hat I\hat
{\times}\hat W^{\dagger} = \hat I,
\eqno(6.15a)
$$
$$\hat A\hat {\times}\hat B\rightarrow \hat W\hat {\times}
(\hat A\hat {\times}\hat B)\hat {\times}\hat W^{\dagger} =
$$
$$
= (\hat W\times \hat T\times A\times \hat T\times \hat
W^\dagger)\times (\hat T\times \hat W^\dagger)^{-1}\times \hat
T\times (\hat W\times
$$
$$
\times
 \hat T)^{-1}\times (\hat W\times \hat
T\times \hat B\times \hat T\times \hat W^\dagger) =
$$
$$
= \hat A'\times (\hat W^{\dagger}\times \hat T\times \hat
W)^{-1}\times \hat B' = \hat A'\times \hat T\times \hat B' =  \hat
A'\hat {\times}\hat B',\; etc.
\eqno(6.15b)
$$
Note that the invariance is ensured by the {\it numerically
invariant values of the isounit and of the isotopic
element under nonunitary-isounitary transforms,}
$\hat I \rightarrow \hat I' \equiv \hat I,\; \; \;  A\hat \times B\rightarrow A'\hat \times' b'
\equiv A'\hat \times B',$
in a way fully equivalent to the invariance of quantum mechanics,
as expected to be necessarily the case due to the preservation of the abstract axioms
under isotopies. The resolution of the catastrophic inconsistencies for noninvariant
theories is then consequential.

Hadronic  mechanics has nowadays clear
experimental verifications in particle physics, nuclear physics, superconductivity,
chemistry, astrophysics, cosmology and biology (see monographs [7,8,9h] for details),
which verifications cannot possibly be reviewed here. We merely mention for subsequent
need   the following realization of the isounit for two particles in conditions of mutual penetration
$$
\hat I = Diag. (n_{11}^2,n_{12}^2, n_{13}^2,n_{14}^2)\times Diag. 9n_{21}^2,
n_{22}^2, n_{23}^2, n_{24}^2)\times
$$
$$
\times e^{N\times (\hat\psi / \psi)\times \int d^3r \times
\psi^\dag_{\downarrow}(r)\times \psi_{\uparrow}(r)}
\eqno(6.16)
$$
where $n_{ak}^2, a = 1, 2, k = 1, 2, 3$ are the semiaxes of the ellipsoids
representing the two particles, $n_{a4}, a = 1, 2$ represents their density,
$\hat \psi$ represents the isowavefunction, $\psi$ represents the conventional
wavefunction (that for $\hat I = 1$), and $N$ is a positive constant. Note the clearly nonlinear, nonlocal-integral and nonpotential character of the interactions represented by isounit (6.16).

The use f the above isounit permitted R. M. Santilli and D. Shillady  to reach
the first exact and invariant  representation 
 of the main characteristics of the hydrogen, water and other molecules, said
representation being achieved directly from first axiomatic principles without  {\it
ad hoc} parameters, or  adulterations via the screenings of the Coulomb law under which the notion of quantum loses any physical or mathematical meaning, thus rendering questionable the very name of 
"quantum chemistry"  (see [7b] for details).
. In reality, due to its
nonunitary structure, {\it hadronic chemistry} contains as particular cases all infinitely
possible screenings of the Coulomb laws.

To understand this results, one should note that quantum mechanics is indeed exact for the structure of {\it one}  hydrogen atoms, but the same mechanics is no longer exact for {\it two} hydrogen atoms combined into the hydrogen molecule due to the historical inability to represent the last $2 \%$ of the binding energy as well as due to other insufficiencies. The resolution of these insufficiencies was achieved by  hadronic chemistry [7b] precisely via isounit (6.16), namely, via the time invariant  representation of the nonlinear, nonlocal and nonpotential interactions occurring in the deep overlapping of the wavepackets of electrons in valence bonds. The new structure models of hadrons presented below is essentially an applications in particle physics of these advances achieved in chemistry.

\vskip0.30cm


\noindent {\large {\bf 7. The new structure model of hadrons with massive physical constituents produced free in spontaneous decays.}   In this section, we show that hadronic mechanics permits the exact and (time) invariant representation of "all"  characteristics of the neutron as a new bound state of a proton and an electron suitable for experimental verifications; we extend the results to the new structure model of all hadrons with massive physical constituents that can be produced free in the spontaneous decays;  we show the compatibility of these advances with the standard model when restricted to provide only the final Mendeleev-type classification of hadrons; and we show the capability of hadronic mechanics as being the sole theory capable of permitting quantitative representations of the possible interplay between  matter and the ether, since the latter requires "positive" binding-like energies for which quantum mechanics admits no physically consistent solutions.
When the Schr\"odinger-Santilli isoequation is worked out in
detail for a bound state of two particles with spin in condition of total mutual penetration, there is the emergence of a strongly {\it repulsive} interaction for triplet couplings (parallel spins), and a strongly {\it attractive} interaction for singlet coupling (antiparallel spin). It should be indicated that, to prevent a prohibitive length, this section is primarily dedicated to the "synthesis" of (unstyable) hadrons, while their spontaneous decays is treated elsewhere.

The case of interest here is the lifting of the  Schr\"odinger equation for a conventional two-body bound state (such as the positronium or the hydrogen atom) via isounit (6.16) where both particles are assumed to be spheres of radius $1 F$ for simplicity. Hence, we consider the simple lifting of quantum bound states characterized by the following Animalu-Santilli isounit [7a]
$$
\hbar = 1 \rightarrow \hat I = U\times U^\dag =  e^{k\times (\psi/\hat \psi)\times \int dr^3\times \psi^\dag_{\downarrow}(r)\times \psi_{\uparrow}(r)},
\eqno(7.1)
$$
where $\psi$ is the wavefunction of the quantum state and $\hat \psi$ is that of the corresponding hadronic state.

In all  cases of singlet coupling the lifting 
yields a strongly attractive {\it Hulten
potential}  that, as well known, behaves at short distances like the Coulomb potential, thus  absorbing the
latter, and resulting in the expressions achieved in the original proposal [5a] (for reviews see [7a,7b,9g])
$$
U\times[\left(\frac{1}{2\mu_1}p_1^{ 2}+\frac{1}{2\mu_2}p_2^{2}
 - V_{Coul}(r_{12}) \right)\times |\psi \rangle]\times U^\dag  \approx
$$
$$
        \approx \left(-\frac{\hbar^2}{2\times \bar m_1}\times\nabla^2_1-
            \frac{\hbar^2}{2\times \bar m_2}\times\nabla^2_2-
            V_{o}\times\frac{e^{-r_{12}\times b}}{1-e^{-r_{12}\times b}}\right)\times |\hat\psi \rangle,
\eqno{(7.2)}
$$
where the original Coulomb interaction has been absorbed by the Hulten constant $V_{o}$ and the liftings $m_k \rightarrow \bar m_k, k = 1, 2$, are new {\it mass isorenormalizations,} that is, renormalizations caused by non-Hamiltonian (or non-Lagrangian) interactions.

Needless to say, the isorenormalization of the mass implies that of the remaining intrinsic characteristics of particles. This assures the departure from quantum mechanics as necessary for the problem at hand.

Detailed studies have shown that the constituents of a bound state described by hadronic mechanics are no longer irreducible representations of the conventional Poincar\'e symmetry (a necessary departure due to the lack of a Keplerian structure indicated earlier), and are characterized instead by irreducible isorepresentations of the Poincar\'e-Santilli isosymmetry [6]. For this reason they are called {\it isoparticles} and are denoted with conventional symbols plus a "hat", such as $\hat e^{\pm}, \hat \pi^{\pm}, \hat p^{\pm},$ etc.

The mechanism permitting physically consistent equations for two-body bound states requiring a "positive" binding-like  energy, as it is the case for the $\pi^o$ and the neutron, is due to the mass isorenormalization since it achieves such an increased value under which the Hulten binding energy can be negative.

In fact, for the case of the $\pi^o$ according to synthesis (7.2) the isorenormalized masses of the individual isoelectrons become of the order of $70 MeV$, while for the case of the synthesis of the neutron according to Eq. (7.2), the isonormalized mass of the electron (assuming that the proton is at rest) acquires a value of the order of $1.39 MeV$, thus allowing a negative binding energy in both cases.

Via the use of hadronic mechanics, the original proposal [5a] achieved already in 1978 a {\it new structure model of $\pi^o$ meson as a compressed positronium, thus identifying the physical constituents with ordinary electron and positrons although in an altered state caused by their condition of total mutual penetration.} This permitted the numerical, exact and invariant representation of {\it all} characteristics of the $\pi^o$ via the following single structural equation}
$$
        U\times H_{positr}\times U^\dag \approx  \left(-\frac{\hbar^2}{2\times \bar m_1}\times\nabla^2_1-
            \frac{\hbar^2}{2\times \bar m_2}\times\nabla^2_2-
            V_{Hult}\times\frac{e^{-r_{12}\times b}}{1-e^{-r_{12}\times b}}\right)\times |\hat\psi \rangle = E\times  |\hat\psi \rangle
\eqno{(7.3a)}
$$
$$
m_k = _e = 0.511 MeV, k = 1, 2, \; \; \; E = 134.97 MeV, \; \; \; \tau = 8.4 x 10^{-17} s, \; \; \; R = b^{-1}  = 10^{-13} cm.
\eqno(7.3b)
$$
where the latter expressions are {\it subsidiary constraints} on the former (see [7a,7b,9g] for reviews).

The above results were extended in the original proposal [5a] to all mesons resulting in this way in a structure model of all mesons with massive physical constituents that can be produced free in the spontaneous decays, generally those with the lowest mode, that are nowadays represented with the symbols
$$
\pi^o = (\hat e^+, \hat e^-)_{HM}, \; \; \; 
\pi^\pm =  (\hat \pi^o, \hat e^\pm)_{HM} = (\hat e^+, \hat e^\pm, \hat e^-)_{HM}, \; \; \; 
 K^o = (\hat \pi^+,
\hat \pi^-)_{HM}, etc.,
\eqno(7.4)
$$
where $e, \pi, K, etc.$ represent conventional {\it particles} as detected in laboratory and $\hat
e, \hat \pi, \hat K, etc.$ represent {\it isoparticles}, namely, the alteration of their intrinsic characteristics (called {\it  mutation} [5a]) caused by their deep mutual penetration.

More technically, conventional particles are charactgerized by unitary irreducible representations of the Poincar\'e symmetry, while isoparticles are characterized by the isounitary irreducible representations of the covering Poincar\'e-Santilli isosymmetry [6].

A few comments are now in order. Firstly, we should note the dramatic departures of the above structure models from  conventional trends in the Mendeleev-type classification of hadrons. To begin, when dealing with classification the emphasis is in searching for "mass spectra." On the contrary, structure model of type (7.4) are known to be {\it  spectra suppressing.} 

In essence, the Hulten potential is known to admit only a finite spectrum of energy level. When all conditions (7.3b) are imposed, the energy levels reduces to only one, that specifically and solely for the meson considered. Needless to say, excited states do exist, but are of {\it quantum} type, that is, whenever the constituents are excited, they exit from the "hadronic horizon" because isounit (7.1) reduces to 1, and quantum mechanics is recovered identically. Consequently, {\it the excited states of structure model (7.3) for the $\pi^o$ are given by the infinite energy levels of the positronium.}

An additional dramatic departure from classification trends is given by the number of constituents. According to the standard model, in the transition from one hadron to another of a given family (such as in the transition from $\pi^o$ to $\pi^+$) the number of quark constituents remain the same. On the contrary, according to hadronic mechanics, the number of constituents must necessarily increase with the increase of the mass, exactly as it is the case for the atomic (nuclear) structure in which the number of constituents  increases in the transition from the H to the He atom (from proton to the deuteron).

The model also achieved a representation of the spontaneous decays with the lowest mode that is generally  interpreted as a tunnel effect of the constituents through the hadronic horizon (rather than the particles being "created" at the time of the decay as requested by the standard model). The remaining decay are the results of rather complex events under non-Hamiltonian interactions still under investigation at this writing.

The representation of  Rutherford's synthesis of the neutron, Eq. (1.1), required considerable additional
studies on the isotopies of angular momentum [6a], spin [6b], Lorentz symmetry [6c], Poincar\'e symmetry [6d], the spinorial covering of the Poincar\'e symmetry [6e], the Minkowskian geometry [6f] and their implications for local realism and all that [6g].

Upon completion of these efforts, Santilli achieved in Ref. [9a] of 1990 the first known, numerically exact and time invariant nonrelativistic representation of {\it all} characteristics of the neutron as a hadronic bound state of a proton assumed to be un-mutated and a mutated electron (or isoelectron)
$$
n = (p^+, \hat e^-)_{HM},
\eqno(7.5)
$$
via the following single structural equation representing the compression of the hydrogen atoms below to the hadronic horizon exactly as originally conceived by Rutherford 
$$
        U\times H_{hydr}\times U^\dag \approx  \left(-\frac{\hbar^2}{2\times \hat m_p}\times\nabla^2_1-
            \frac{\hbar^2}{2\times \bar m_e}\times\nabla^2_2-
            V_{Hult}\times\frac{e^{-r_{12}\times b}}{1-e^{-r_{12}\times b}}\right)\times |\hat\psi \rangle = E\times  |\hat\psi \rangle
\eqno{(7.6a)}
$$
$$
\mu _e = 0.511 MeV, , \; \; \; \mu_p = 938, 27 MeV, \; \; \; E = 939.56 MeV, \; \; \; 
\tau = 886 s, \; \; \; R = 10^{-13} cm.
\eqno(7.6b)
$$
The relativistic extension of the above model was reached in Ref. [9b] of 1993 (see also [6e]) via the isotopies of Dirac's equation, and cannot be reviewed here to avoid a prohibitive length.

Remarkably, despite the disparities between Eqs. (7.3) and (7.6), the Hulten potential admitted again one single energy level, that of the neutron. Under excitation, the isoelectron exits the hadronic horizon (again, because the integral in Eq. (7.1) becomes null) and one recovers the quantum description. Consequently, according to structure model (7.6), {\it the excited states of the neutron are the infinite energy levels of the hydrogen atom.}

The representation of the spin 1/2 of the neutron turned out to be much simpler than expected, as outlined in Figure  4. In particular, {\it the hadronic representation of the synthesis of the neutron does not require any neutrino at all, exactly as originally conceived by Rutherford,} of course, not at the quantum level, but at the covering hadronic level.

Once compressed inside the
proton, in order to have an attractive bond, the electron is constrained to have its spin antiparallel to that of the
proton and, in order to achieve a stable state, the electron orbital momentum  is constrained to coincide with the spin 1/2 of the proton (Figure 4), 
resulting in the following representation  of the spin of the neutron
$$
s_n^{spin} = s_p^{spin} + s_{\hat e}^{spin} + s_{\hat e}^{orb} = {1\over 2} - {1\over 2} + {1\over 2},
\eqno(7.7)
$$
namely, {it the total angular momentum of the isoelectron is null},
$$
s_{\hat e}^{tot} = s_{\hat e}^{spin} + s_{\hat e}^{orb} = 0,
\eqno(7.8)
$$ 
and {\it the spin of the neutron coincides with that of the proton.}

It should be noted that a fractional value of the angular momentum is anathema for quantum mechanics, namely, when defined over a conventional Hilbert space $\cal H$ over the field of complex numbers $C$
(because it causes a departure from its nonunitary structure with a host of problems), while
the same fractional value is fully admissible for hadronic mechanics, namely when defined on an iso-Hilbert space ${\cal{\hat H}}$ over an isofield ${\cal {\hat C}}$ in view of its isounitary structure.

As a simple example, under the isounit and isotopic elements $\hat I = {1\over 2}, \hat T = 2$ and isonormalization $<\hat \psi| \times \hat T \times |\hat \psi> = 1$ the half-off-integer angular momentum $\hat J_3 = {1\over 2}$ admits the isoexpectation  value 1,
$$
<\hat J_3> = {<\hat \psi |\hat \times \hat J_3\hat\times |\hat \psi>\over <\hat \psi |\hat
\times |\hat \psi>} = 1.
\eqno(7.9)
$$

The above occurrence should not be surprising for the reader familiar with   hadronic mechanics. In fact, the sole admission of conventional values of the angular momentum would imply the admission of the perpetual motion for, say, an electron orbiting in the core of a star. In  the transition from motion in a quantized orbit in vacuum in an atomic structure to motion within the core of a star, the angular momentum assumes an arbitrary, locally varying value. The only reason for the orbital value ${1\over 2}$ for the neutron is the existence of the constraint restricting the angular momentum of the isoelectron to coincide with the spin of the proton (Figure 4).



\begin{figure}[htbp] 
   \centering
   \includegraphics[width=2.5in]{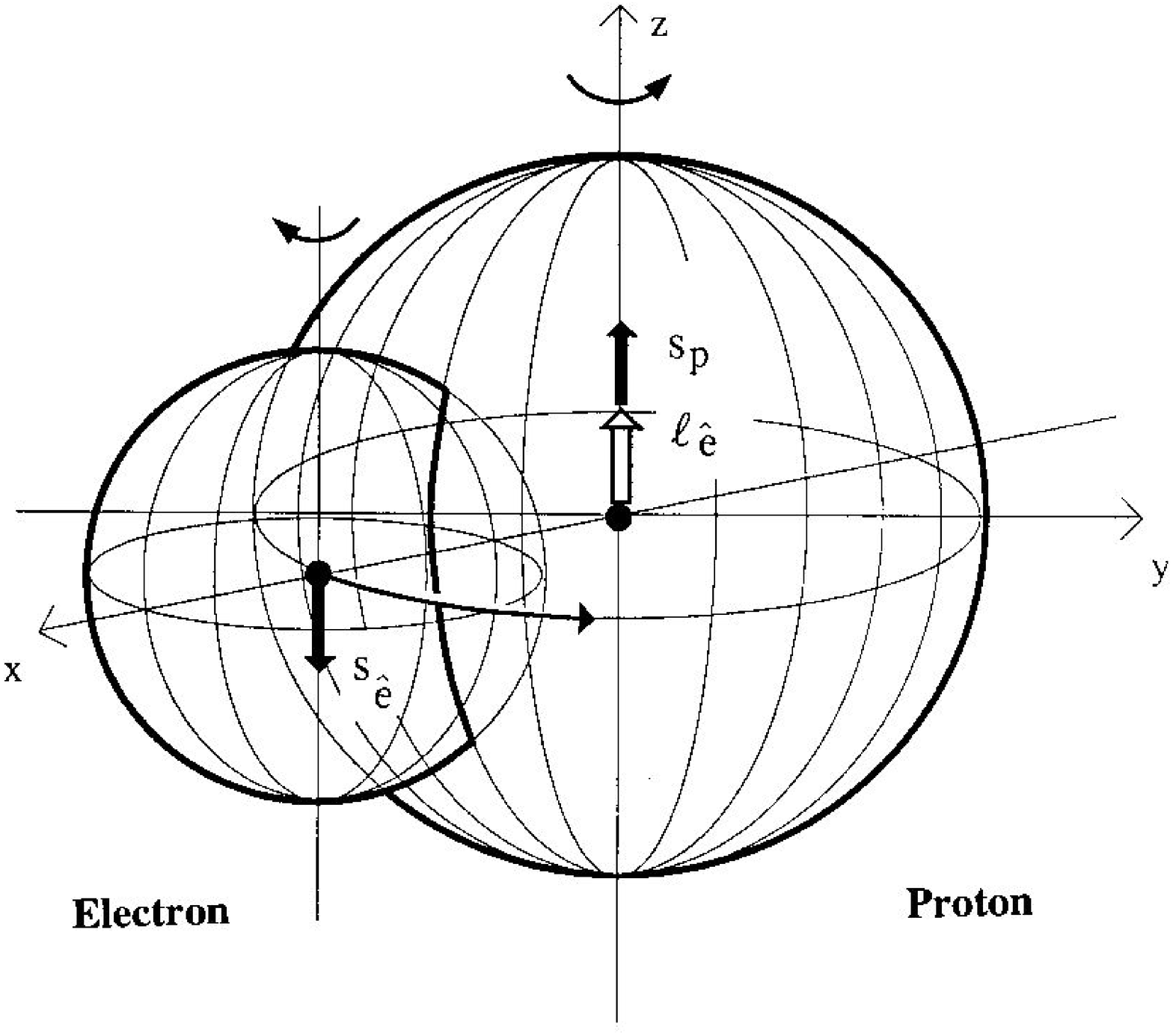} 
 \centering   \caption{{\it . A schematic view of the intrinsic and orbital angular momenta in Rutherford's synthesis of the neutron according to hadronic mechanics.}}
\end{figure}


The nonrelativistic, exact and invariant representation of the anomalous magnetic moment of the neutron ($-1.913 m_N$) from those of the proton and of the electron was also achieved for the first time by R. M. Santilli in Ref. [9a] of 1990.

The magnetic moment of Rutherford's neutron is characterized by three contributions, the magnetic moment of the proton, that of the isoelectron, and that caused by the orbital motion of the isoelectron. Note that for quantum mechanics the third contribution is completely missing because all particles are considered as points, in which case the electron cannot rotate inside the proton. As well known, the inability by quantum mechanics to treat the orbital motion of the electron inside the proton (due to its point-like character) was the very origin of the conjecture of the neutrino.

With reference to the orientation of Figure 4, and by keeping in mind that a change of the sign of the charge implies a reversal of the sign of the magnetic moment, the representation of Ref. (9a) is based on the identity
$$
\mu_n = \mu_p + \mu_{\hat e-intrinsic} - m_{\hat e-orbital} = -1.9123 \mu_N,
\eqno(7.10)
$$

Since the spin of the proton and of the electron can be assumed to be conventional in first approximation, we can assume that their intrinsic magnetic moments  are conventional, i.e.,
$$
\mu_p  = + 2.793 m_N, \; \; \; 
\mu_{\hat e - intrinsic} = \mu_e = - 1.001 \mu_B = - 1,837.987 \mu_N,
\eqno(7.11)
$$
consequently
$$
\mu_p + \mu_e = 1,835 \mu_N,
\eqno(7.12)
$$

It is then evident that the anomalous magnetic moment of the neutron originates from the magnetic moment of the orbital motion of the isoelectron inside the proton, namely, a contribution that has been ignored since Rutherford's time until treated in Ref. [9a].

It is easy to see that the desired exact and invariant representation of the anomalous magnetic moment of the neutron is characterized by the following numerical values
$$
\mu_{\hat e - orbital} = +1.004 \mu_B, \; \; \; 
\mu_{\hat e - total} = 3\times 10^{-3} \mu_B, \; \; \; 
\mu_n = -1.9123 \mu_N,
\eqno(7.13)
$$
and this completes our nonrelativistic review. Note that the small value of the total magnetic moment of the isoelectron is fully in line with the small value of its total angular momentum (that is null in first approximation due to the assumed lack of mutation of the proton).

We regret to be unable to review the numerically exact and time invariant {\it relativistic} representation of the anomalous magnetic moment of the neutron [6e,9b] because it provides much deeper insights than the preceding one with particular reference to the {\it density} of hadrons, $n_4^2$, in isounit (6.16), that is completely absent in conventional treatment via the standard model. In fact, the numerical value of $n_4$ obtained from the fit of the data on the fireball of the Bose-Einstein correlation permits the exact representation of the neutron anomalous magnetic moment without any additional quantity or unknown parameters.

As one can see, the spontaneous decay of the neutron into physical, actually observed particles is given by
$$
n = (p^+, \hat e^-)_{HM} \rightarrow p^+ + e^-,
\eqno(7.14)
$$
as a mere tunnel effect of the massive constituents through the hadronic horizon, after which particles reacquire their conventional quantum characteristics. Assuming that the neutron and the proton are {\it
isolated} and at rest in the spontaneous decay, the electron is emitted with $0.782 MeV$ energy. When the decaying neutron is a member of a nuclear structure, the energy possessed by the electron is generally less than $0.782 MeV$ and varies depending on the angle of emission, as indicated in Section 2.

The behavior of the angular momentum for reaction (7.14) can be interpreted at the level of hadronic mechanics and related Poincar\'e-Santilli isosymmetry via the transformation of the orbital into linear motions without any need for the neutrino, in the same way as no neutrino is needed for the neutron synthesis .

The extension of the model to all baryons resulted to be elementary, with models of the type
$$
n = (\hat p^+, \hat e^-)_{HM} \approx  (p^+, \hat e^-)_{HM}, \; \; \; \Lambda =  (\hat p^+, \hat \pi^-)_{HM} \approx  (\hat n, \hat \pi^o)_{HM},
\eqno(7.15a)
$$
$$ \Sigma^+ =  (\hat p^+, \hat \pi^o)_{HM} \approx  (\hat {\bar n}, \hat \pi^+)_{HM}, \; \; \; \Sigma^o =  (\hat \Lambda, \hat e^+, \hat e^-)_{HM}, \; \; \; \Sigma^- =  (\hat n, \hat \pi^-)_{HM} \approx  (\hat {\bar p}^-, \hat \pi^o)_{HM}, \; \; \; etc.
\eqno(7.15)
$$
where one should note the  equivalence of seemingly different structure models due to the indicated mutation of the constituents.

Compatibility of the hadronic structure models with the SU(3)-color Mendeleev-type
classification was first suggested in Ref. [5d] and resulted to be possible in a variety of ways, such as, via a multivalued hyperunit [7a] consisting of a set of isounits each characterizing the structure of one individual hadrons in a given unitary multiplet 
$$
\hat I = Diag. (\hat I_{\pi^o}, \hat I_{\pi^+}, \hat I_{\pi^-}, \hat I_{K^o_S}, ... ) = U\times U^\dag > 0.
\eqno(7.16)
$$
The lifting of SU(3)-color symmetries under the above hyperunit is isomorphic to the conventional symmetry due to the positive-definiteness of the hyperunit,
$$
U\times SU(3)\times U^dag \approx SU(3),
\eqno(7.17)
$$
thus ensuring the preservation of all numerical results of the Mendeleev-type classifications due to the preservation of the structure constants, Eqs. (6.12e).

In closing, the reader may have noted the dichotomy between  the etherino hypothesis of Section 5 for the synthesis of the neutron and its absence in this section for the same topic. This is due to the fact that the lifting $cal H\rightarrow \hat{\cal{H}}$ 
 implicitly represents the absorption of the needed spin and energy from the ether or from the environment (such as the interior of a star), thus clarifying that, unlike the neutrino, the etherino {\it is not} a physical particle in our spacetime, but merely represents the indicate {\it transfer} of features. We therefore have the following equivalence
$$
n = (p^+, a^o, e^-)_{QM} \approx (\hat p^+, \hat e^-)_{HM},
\eqno(7.18)
$$
with the understanding that the Schr\"odinger equation is physically inconsistent for the QM formulation, while its isotopic image is fully consistent.

We can therefore conclude by saying that {\it hadronic mechanics is the first and only theory known to the author for quantitative invariant studies of the interplay between matter and the background medium, whether the ether or the hyperdense medium inside hadrons.} 
\vskip0.30cm


\noindent {\large {\bf 8. New clean energies permitted by the absence of neutrinos and quarks.}  Molecular, atomic and nuclear structures have provided immense benefits to mankind because their
constituents can be produced free. Quark theories on the structure of hadrons have no practical value
whatever  because, by comparison, quarks by conception cannot be produced free. 

On the contrary, the structure model of hadrons with physical constituents that can be produced free allows the prediction of new clean energies originating in the structure of individual hadrons, rather than in nuclei, todays known as {\it hadronic energies} [9c], that could provide the first  industrial application of hadron physics.

In fact, {\it the neutron is the biggest reservoir of clean energy available to
mankind} because: 1) The neutron is naturally unstable; 2) When decaying it releases a large amount of
energy  carried by the emitted electron; and 3) Such energy can be 
easily trapped with a thin metal shield.

Moreover, hadronic energy  is  two-fold because, when the decay of the
neutron occurs in a conductor, the latter acquires a positive charge while the shield
trapping the electron acquires a negative charge, resulting in a new clean production of continuous current originating in the structure of the neutron 
first proposed by Santilli in Ref. [9c] and today known as {\it hadronic battery}  [8]. The second
source of energy is thermal and it is given by the  heat acquired by the  shield trapping the emitted
electrons.

Recall that, unlike the proton, the neutron is {\it naturally unstable.} Consequently, it must admit a
{\it stimulated decay.} That predicted by hadronic mechanics was first proposed by Santilli [9c] in  1994
and it is given by hitting a selected number of nuclear isotopes, called {\it hadronic fuels,} with hard
photons $\gamma_{res}$ having energy (frequency) given by a submultiple of the difference of energy between the neutron and the proton
$$
\Delta E = m_n - m_p = 1.293 MeV = h\times \nu_{reson}, 
\eqno(8.1a)
$$
$$
E_{res} = {1.294\over n} MeV, n = 1, 2, 3, ...,
\eqno(8.1b)
$$



\begin{figure}[htbp] 
   \centering
   \includegraphics[width=4in]{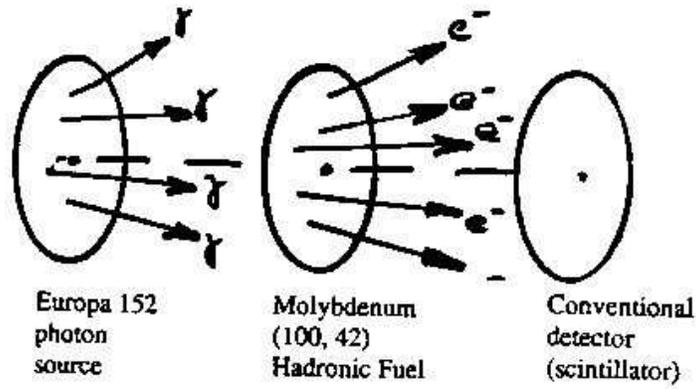} 
 \centering   \caption{{\it  The view illustrates a 
"hadronic fuel",  the MO(100,42), that, when hit by a neutron resonating frequency, is predicted to experience a
stimulated decay into an unstable isotope that, in turn, decays spontaneous into a final stable isotope
with the total emission of two highly energetic electrons, thus realizing the conditions of Figure 5.}}
\end{figure}


under which the isoelectron is predicted to be excited and consequently cross the $1 F$ hadronic horizon, resulting in the stimulated decay 
$$
\gamma_{res} + n \rightarrow p^+ + e^-.
\eqno(8.2)
$$

The energy gain is beyond scientific doubt, because the use of 1/10-th of the exact resonating
frequency (8.1b) could produce clean energy  up to 100-times the original value, depending on the energy of the released betas. Note that the energy of photons not causing stimulated decay is not lost, because absorbed by the hadronic fuel, thus being part of the heat balance.

One among numerous cases of hadronic energy proposed for test in Ref. [5c] is given by
$$
\gamma_{res} + Mo(100, 42)\rightarrow Tc(100, 43) + \beta^-,
\eqno(8.3a)
$$
$$
Tc(100, 43)\rightarrow Ru(100, 44) + \beta^-,
\eqno(8.3b)
$$
where the first beta decay is stimulated while the second is natural and occurs in 18 sec. 

Note that the conventional nuclear energy is based on the disintegration of {\it large and heavy nuclei, } thus causing well known dangerous radiations and leaving dangerous waste. By comparison,  hadronic energy is based on the use of {\it light nuclei} as in thecase of Eqs. (8.3), thus releasing no harmful radiation and leaving no harmful waste because both the original nucleus Mo(100, 42) and the final one Ru(100,44) are natural, light and stable elements (for additional studies, see [9g]).

As proposed in Ref. [9d], the above stimulated decay of the neutron could be of assistance to conventional nuclear energy since it would allow the stimulated decay of radioactive nuclear waste  via the use of a coherent beam of resonating gammas, plus additional feature, such as high intensity electric fields. These conditions would cause a sudden increase ofd positive charges plus an ellipsoidal deformation of large nuclei under which their decay is unavoidable. The latter equipment is predicted to be sufficiently small to be used by nuclear power plants in their existing pools, thus rendering conventional nuclear energy more accepted by society.

\vskip0.30cm


\noindent {\large {\bf 9. The much needed experimental resolutions.}} The international physics community has spent to date in neutrino and quark conjectures well in excess of ten billion dollars of public funds from various countries while multiplying, rather than resolving the controversies as indicated in Sections 2 and 2. It is evident that the physics community simply cannot continue this trend without risking a historical condemnation by posterity. When possible new energies so much needed by mankind emerge from alternative theories without neutrinos and quarks as physical particles in our spacetime, the gravity of the case emerge in its full light.



\begin{figure}[htbp] 
   \centering
   \includegraphics[width=3in]{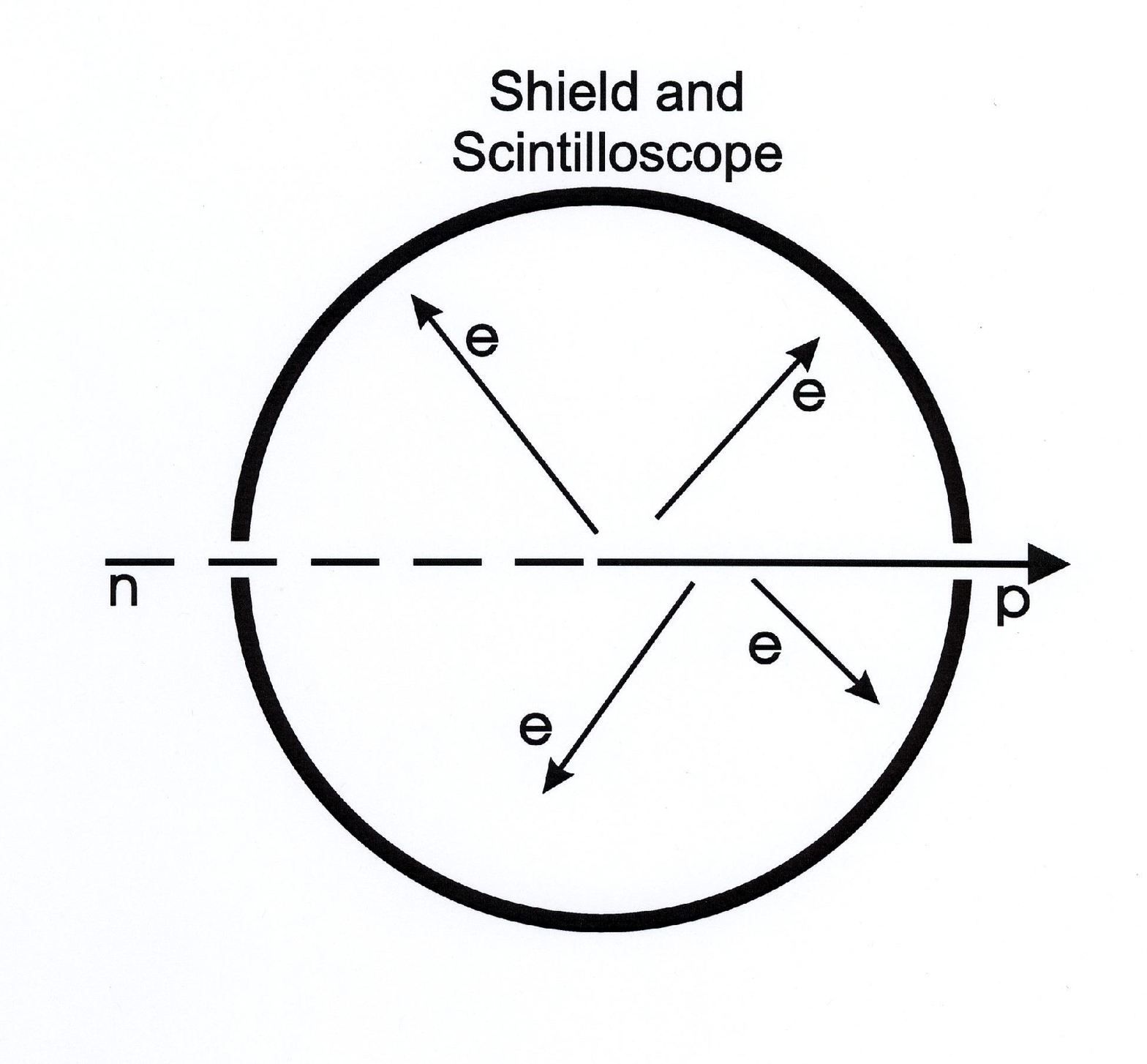} 
 \centering   \caption{{\it  A schematic view of the proposed measurement of the energy of the electron in
spontaneous neutron decays to ascertain whether there is any energy left for the neutrino.
}}
\end{figure}


 In the hope of contributing toward the much 
needed expriental resolutions, in this paper we propose the
following basic tests.

{\bf Proposed First Experiment: to resolve whether or not there is energy
available for the neutrino in the neutron decay.} This experiment 
can be done today in numerous ways. That recommended in this
note, apparently for the first time, is to conduct systematic
 measurements of the energy of the electron emitted in the decay
of a coherent beam of {\it low energy} (e.g., thermal) neutrons as depicted in Fig. 6. 

The detection of energies of the electrons systematically {\it less} than  $0.78 MeV$ (plus the neutron energy), would eliminate the third hypothesis on tyhe neutron decay, Eq. (5.5), and support the firsty and second hypotheses, Eqs. (5.3) and (5.4), but would be unable to distinguish between them.

The detection of electron energy systematically given by $0.78 MeV$ (plus the neutron energy) would disprove the emission of a neutrino or an etherino in neutron decays, and support the third hypothesis, Eq. (5.) in clear favbor ofg the continuous creation of matter in the unioverse.

Note that the conduction of  the proposed test with ``high energy" neutrons would not be resolutory   because the {\it variation} of the electron energy expected to be absorbed by the neutrino would be excessively smaller than   the electron energy.   

The conduction of the test via nuclear beta decays is also not recommendable due to the indicated expected dependence of the electron energy from the direction of beta emission, which dependence is ignorable for the case of decay of individual neutrons. 

The author has been unable to identify in the literature any conduction of the proposed trest since all available experiments refer to {\it nuclear} beta decays rather than that of individual {\it neutrons}. Any indication by interfested colleagues of specific reference to tests similar to that herein proposed would be gratefully appreciatred.



\begin{figure}[htbp] 
   \centering
   \includegraphics[width=3in]{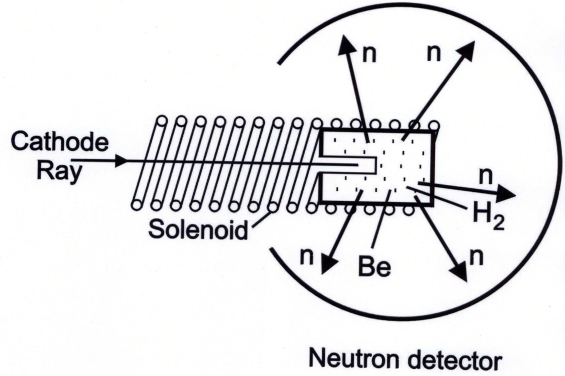} 
 \centering   \caption{{\it  A schematic view of the proposed laboratory synthesis of the 
neutron from protons and electrons to identify
the needed energy.
}}
\end{figure}


I
{\bf Proposed Second Experiment: Achieve the laboratory synthesis of 
the neutron to identify the needed energy.}
The  first attempt at synthesizing the neutron in laboratory 
known to this author was conducted with positive outcome in Brazil by the Italian
priest-physicist Don Borghi and his associates [9e] (see [3b] for a review). The  tests were apparently successful, although the experimental set up does not allow the measurement of the{\it  energy} needed for the synthesis.

The latter information can be obtained nowadays in a variety of ways. That recommended in this note,   consists in sending a coherent electron beam  against a beryllium mass saturated with hydrogen and kept at low temperature (so  that the protons of the hydrogen atoms can be approximately
considered to be at rest). A necessary condition for credibility  is that said protons and electrons be
polarized to have antiparallel spins (singlet couplings), because large repulsions are predicted for triplet couplings at very short distances for particles with spin, as it is the case for the coupling of ordinary gears. Since the proton and the electron have opposite charges, said polarization can be achieved with the same  magnetic field as illustrated in Fig. 7.

Neutrons that can possibly be synthesized in this way will escape from the beryllium mass and can be detected  with standard means. The detection of neutron produced with electron kinetic  energies systematically  in excess of $0.78 MeV$ would confirm the neutrino hypothesis. The systematic detection of neutrons  synthesized either at the threshold energy of $0.78 MeV$ or less would support alternative hypotheses, such as that of the etherino, and render polausible the hypothesis of continuous  creation of matter in the universe via the neutron synthesis as studied in Section 5.

{\bf Proposed Third Experiment: Test the stimulated decay of the 
neutron as a source for new clean energies.}  
The test of the  
stimulated decay of the neutron proposed in Ref. [9c], Eqs. (8.3) and Figure 5,  
was successfully conducted by N. Tsagas and his associates [9f] (see Ref. [9g for a review and upgrading).  As illustrated in Fig. 6, the latter experiment was conducted via a disk 
of Europa 152 (emitting photons precisely with the needed resonating frequency)
 coupled to a disc of  molybdenum, 
the pair being contained
inside a scintilloscope for the detection of the emitted electrons, 
the experimental set up being
suitably shielded, as customary. The test was successful because it detected 
electrons emitted by the indicated pair with energy
bigger than $1\, MeV$  since  
electrons from the Compton scattering of photons and
atomic electrons can at most have $1\, MeV$. the same test can be repeated in a variety of way with different hadronic fuels (see [9c] for alternatives).

Needless to say, despite their positive outcome, the available 
results on the proposed three tests are
empirical, rudimentary and inconclusive. Rather than being reasons for 
dismissal, these insufficiencies establish instead the need for the finalization 
of the proposed basic experiments, 
also in view of their rather large scientific  and societal relevance.

\vskip0.30cm

\noindent {\large {\bf Acknowledgments.} The author has no word to express his appreciation to IARD members because without their serious commitment to knowledge this paper could not have been discussed during the meeting of 2006, let alone published. Additional thanks for invaluable comments and criticisms are due to H. E. Wilhelm of the Institute for Basic Research, Palm Harbor, Florida, and L. Daddi  of the Naval Academy, Livorno, Italy.

\end{document}